
\message
{JNL.TEX version 0.92 as of 6/9/87.  Report bugs and problems to Doug Eardley.}
\message
{This is X.G. Wen's copy}

\catcode`@=11
\expandafter\ifx\csname inp@t\endcsname\relax\let\inp@t=\input
\def\input#1 {\expandafter\ifx\csname #1IsLoaded\endcsname\relax
\inp@t#1%
\expandafter\def\csname #1IsLoaded\endcsname{(#1 was previously loaded)}
\else\message{\csname #1IsLoaded\endcsname}\fi}\fi
\catcode`@=12



\font\twelverm=amr10 scaled 1200    \font\twelvei=ammi10 scaled 1200
\font\twelvesy=amsy10 scaled 1200   \font\twelveex=amex10 scaled 1200
\font\twelvebf=ambx10 scaled 1200   \font\twelvesl=amsl10 scaled 1200
\font\twelvett=amtt10 scaled 1200   \font\twelveit=amti10 scaled 1200
\font\twelvesc=amcsc10 scaled 1200  \font\twelvesf=amssmc10 scaled 1200
\skewchar\twelvei='177   \skewchar\twelvesy='60


\def\twelvepoint{\normalbaselineskip=12.4pt plus 0.1pt minus 0.1pt
  \abovedisplayskip 12.4pt plus 3pt minus 9pt
  \belowdisplayskip 12.4pt plus 3pt minus 9pt
  \abovedisplayshortskip 0pt plus 3pt
  \belowdisplayshortskip 7.2pt plus 3pt minus 4pt
  \smallskipamount=3.6pt plus1.2pt minus1.2pt
  \medskipamount=7.2pt plus2.4pt minus2.4pt
  \bigskipamount=14.4pt plus4.8pt minus4.8pt
  \def\rm{\fam0\twelverm}          \def\it{\fam\itfam\twelveit}%
  \def\sl{\fam\slfam\twelvesl}     \def\bf{\fam\bffam\twelvebf}%
  \def\mit{\fam 1}                 \def\cal{\fam 2}%
  \def\sc{\twelvesc}		   \def\tt{\twelvett}
  \def\sf{\twelvesf}
  \textfont0=\twelverm   \scriptfont0=\tenrm   \scriptscriptfont0=\sevenrm
  \textfont1=\twelvei    \scriptfont1=\teni    \scriptscriptfont1=\seveni
  \textfont2=\twelvesy   \scriptfont2=\tensy   \scriptscriptfont2=\sevensy
  \textfont3=\twelveex   \scriptfont3=\twelveex  \scriptscriptfont3=\twelveex
  \textfont\itfam=\twelveit
  \textfont\slfam=\twelvesl
  \textfont\bffam=\twelvebf \scriptfont\bffam=\tenbf
  \scriptscriptfont\bffam=\sevenbf
  \normalbaselines\rm}



\def\beginlinemode{\endmode
  \begingroup\parskip=0pt \obeylines\def\\{\par}\def\endmode{\par\endgroup}}
\def\beginparmode{\endmode
  \begingroup \def\endmode{\par\endgroup}}
\let\endmode=\par
{\obeylines\gdef\
{}}
\def\singlespace{\baselineskip=\normalbaselineskip}

\def\oneandahalfspace{\baselineskip=\normalbaselineskip
  \multiply\baselineskip by 3 \divide\baselineskip by 2}
\def\doublespace{\baselineskip=\normalbaselineskip \multiply\baselineskip by 2}

\newcount\firstpageno
\firstpageno=2
\footline={\ifnum\pageno<\firstpageno{\hfil}\else{\hfil\twelverm\folio\hfil}\fi}
\def\toppageno{\global\footline={\hfil}\global\headline
  ={\ifnum\pageno<\firstpageno{\hfil}\else{\hfil\twelverm\folio\hfil}\fi}}
\let\rawfootnote=\footnote		
\def\footnote#1#2{{\rm\singlespace\parindent=0pt\parskip=0pt
  \rawfootnote{#1}{#2\hfill\vrule height 0pt depth 6pt width 0pt}}}
\def\raggedcenter{\leftskip=4em plus 12em \rightskip=\leftskip
  \parindent=0pt \parfillskip=0pt \spaceskip=.3333em \xspaceskip=.5em
  \pretolerance=9999 \tolerance=9999
  \hyphenpenalty=9999 \exhyphenpenalty=9999 }
\def\dateline{\rightline{\ifcase\month\or
  January\or February\or March\or April\or May\or June\or
  July\or August\or September\or October\or November\or December\fi
  \space\number\year}}
\def\received{\vskip 3pt plus 0.2fill
 \centerline{\sl (Received\space\ifcase\month\or
  January\or February\or March\or April\or May\or June\or
  July\or August\or September\or October\or November\or December\fi
  \qquad, \number\year)}}


\hsize=6.5truein
\hoffset=0truein
\vsize=8.9truein
\voffset=0truein
\parskip=\medskipamount
\def\\{\cr}
\twelvepoint		
\doublespace		
\overfullrule=0pt	




\def\title			
  {\null\vskip 3pt plus 0.2fill
   \beginlinemode \doublespace \raggedcenter \bf}

\def\author			
  {\vskip 3pt plus 0.2fill \beginlinemode
   \singlespace \raggedcenter\sc}

\def\affil			
  {\vskip 3pt plus 0.1fill \beginlinemode
   \oneandahalfspace \raggedcenter \sl}

\def\abstract			
  {\vskip 3pt plus 0.3fill \beginparmode
   \oneandahalfspace ABSTRACT: }

\def\endtitlepage		
  {\endpage			
   \body}

\def\body			
  {\beginparmode}		

\def\head#1{			
  \goodbreak\vskip 0.5truein	
  {\immediate\write16{#1}
   \raggedcenter \uppercase{#1}\par}
   \nobreak\vskip 0.25truein\nobreak}

\def\beginitems{
\par\medskip\bgroup\def\i##1 {\item{##1}}\def\ii##1 {\itemitem{##1}}
\leftskip=36pt\parskip=0pt}
\def\enditems{\par\egroup}

\def\beneathrel#1\under#2{\mathrel{\mathop{#2}\limits_{#1}}}

\def\refto#1{$^{#1}$}		

\def\references			
  {\head{References}		
   \beginparmode
   \frenchspacing \parindent=0pt \leftskip=1truecm
   \parskip=8pt plus 3pt \everypar{\hangindent=\parindent}}

\def\referencesnohead   	
  {                     	
   \beginparmode
   \frenchspacing \parindent=0pt \leftskip=1truecm
   \parskip=8pt plus 3pt \everypar{\hangindent=\parindent}}

\gdef\refis#1{\item{#1.\ }}			

\gdef\journal#1, #2, #3, 1#4#5#6{		
    {\sl #1~}{\bf #2}, #3 (1#4#5#6)}		

\def\endreferences{\body}

\def\figurecaptions		
  {\endpage
   \beginparmode
   \head{Figure Captions}
}

\def\endpage			
  {\vfill\eject}

\def\endpaper			
  {\endmode\vfill\supereject}


\def\heading				
  {\vskip 0.5truein plus 0.1truein	
   \beginparmode \def\\{\par} \parskip=0pt \singlespace \raggedcenter}

\def\subheading				
  {\vskip 0.25truein plus 0.1truein	
   \beginlinemode \singlespace \parskip=0pt \def\\{\par}\raggedcenter}

\def\tag#1$${\eqno(#1)$$}

\def\align#1$${\eqalign{#1}$$}

\def\aligntag#1$${\gdef\tag##1\\{&(##1)\cr}\eqalignno{#1\\}$$
  \gdef\tag##1$${\eqno(##1)$$}}

\def\endaligntag{}

\def\overset #1\to#2{{\mathop{#2}\limits^{#1}}}
\def\underset#1\to#2{{\let\next=#1\mathpalette\undersetpalette#2}}
\def\undersetpalette#1#2{\vtop{\baselineskip0pt
\ialign{$\mathsurround=0pt #1\hfil##\hfil$\crcr#2\crcr\next\crcr}}}


\def\ref#1{Ref.~#1}			
\def\Ref#1{Ref.~#1}			
\def\[#1]{[\cite{#1}]}
\def\cite#1{{#1}}
\def\(#1){(\call{#1})}
\def\call#1{{#1}}
\def\taghead#1{}
\def\frac#1#2{{#1 \over #2}}

\def\12{{1\over2}}

\def\sla{\raise.15ex\hbox{$/$}\kern-.57em}
\def\leaderfill{\leaders\hbox to 1em{\hss.\hss}\hfill}
\def\twiddle{\lower.9ex\rlap{$\kern-.1em\scriptstyle\sim$}}
\def\bigtwiddle{\lower1.ex\rlap{$\sim$}}
\def\gtwid{\mathrel{\raise.3ex\hbox{$>$\kern-.75em\lower1ex\hbox{$\sim$}}}}
\def\ltwid{\mathrel{\raise.3ex\hbox{$<$\kern-.75em\lower1ex\hbox{$\sim$}}}}
\def\square{\kern1pt\vbox{\hrule height 1.2pt\hbox{\vrule width 1.2pt\hskip 3pt
   \vbox{\vskip 6pt}\hskip 3pt\vrule width 0.6pt}\hrule height 0.6pt}\kern1pt}
\def\tdot#1{\mathord{\mathop{#1}\limits^{\kern2pt\ldots}}}

\def\pmb#1{\setbox0=\hbox{#1}%
  \kern-.025em\copy0\kern-\wd0
  \kern  .05em\copy0\kern-\wd0
  \kern-.025em\raise.0433em\box0 }

\def\3he{{$^3${\rm He}}}

\def\]{\right]}
\def\[{\left[}

\def\>{\rangle}
\def\<{\langle}
\def\o{\over}

\def\slD{\raise.15ex\hbox{$/$}\kern-.57em\hbox{$D$}}
\def\dsl{\raise.15ex\hbox{$/$}\kern-.57em\hbox{$\Delta$}}
\def\slp{{\raise.15ex\hbox{$/$}\kern-.57em\hbox{$\partial$}}}
\def\nsl{\raise.15ex\hbox{$/$}\kern-.57em\hbox{$\nabla$}}
\def\sla{\raise.15ex\hbox{$/$}\kern-.57em\hbox{$\rightarrow$}}
\def\slla{\raise.15ex\hbox{$/$}\kern-.57em\hbox{$\lambda$}}
\def\slb{\raise.15ex\hbox{$/$}\kern-.57em\hbox{$b$}}
\def\lnp{\raise.15ex\hbox{$/$}\kern-.57em\hbox{$p$}}
\def\lnk{\raise.15ex\hbox{$/$}\kern-.57em\hbox{$k$}}
\def\lnK{\raise.15ex\hbox{$/$}\kern-.57em\hbox{$K$}}
\def\lnq{\raise.15ex\hbox{$/$}\kern-.57em\hbox{$q$}}


\def\pmb#1{\setbox0=\hbox{$#1$}%
\kern-.025em\copy0\kern-\wd0
\kern.05em\copy0\kern-\wd0
\kern-.025em\raise.0433em\box0 }

\def\q2{{Q^2}}
\def\gtwid{\raise.3ex\hbox{$>$\kern-.75em\lower1ex\hbox{$\sim$}}}
\def\ltwid{\raise.3ex\hbox{$<$\kern-.75em\lower1ex\hbox{$\sim$}}}
\def\12{{1\over2}}
\def\part{\partial}

\def\low#1{\lower.5ex\hbox{${}_#1$}}

\def\psl{\raise.15ex\hbox{$/$}\kern-.57em\hbox{$\partial$}}
\def\partt{\raise.15ex\hbox{$\widetilde$}{\kern-.37em\hbox{$\partial$}}}

\def\abs{
         \vskip 3pt plus 0.3fill\beginparmode
         \doublespace ABSTRACT:\ }

\def\topppageno1{\global\footline={\hfil}\global\headline
={\ifnum\pageno<\firstpageno{\hfil}\else{\hss\twelverm --\ \folio
\ --\hss}\fi}}

\def\toppageno2{\global\footline={\hfil}\global\headline
={\ifnum\pageno<\firstpageno{\hfil}\else{\rightline{\hfill\hfill
\twelverm \ \folio
\ \hss}}\fi}}

\catcode`@=11
\newcount\r@fcount \r@fcount=0
\newcount\r@fcurr
\immediate\newwrite\reffile
\newif\ifr@ffile\r@ffilefalse
\def\w@rnwrite#1{\ifr@ffile\immediate\write\reffile{#1}\fi\message{#1}}

\def\writer@f#1>>{}
\def\referencefile{
  \r@ffiletrue\immediate\openout\reffile=\jobname.ref%
  \def\writer@f##1>>{\ifr@ffile\immediate\write\reffile%
    {\noexpand\refis{##1} = \csname r@fnum##1\endcsname = %
     \expandafter\expandafter\expandafter\strip@t\expandafter%
     \meaning\csname r@ftext\csname r@fnum##1\endcsname\endcsname}\fi}%
  \def\strip@t##1>>{}}

\def\citeall#1{\xdef#1##1{#1{\noexpand\cite{##1}}}}
\def\cite#1{\each@rg\citer@nge{#1}}	

\def\each@rg#1#2{{\let\thecsname=#1\expandafter\first@rg#2,\end,}}
\def\first@rg#1,{\thecsname{#1}\apply@rg}	
\def\apply@rg#1,{\ifx\end#1\let\next=\relax
\else,\thecsname{#1}\let\next=\apply@rg\fi\next}

\def\citer@nge#1{\citedor@nge#1-\end-}	
\def\citer@ngeat#1\end-{#1}
\def\citedor@nge#1-#2-{\ifx\end#2\r@featspace#1 
  \else\citel@@p{#1}{#2}\citer@ngeat\fi}	
\def\citel@@p#1#2{\ifnum#1>#2{\errmessage{Reference range #1-#2\space is bad.}%
    \errhelp{If you cite a series of references by the notation M-N, then M and
    N must be integers, and N must be greater than or equal to M.}}\else%
 {\count0=#1\count1=#2\advance\count1
by1\relax\expandafter\r@fcite\the\count0,%
  \loop\advance\count0 by1\relax
    \ifnum\count0<\count1,\expandafter\r@fcite\the\count0,%
  \repeat}\fi}

\def\r@featspace#1#2 {\r@fcite#1#2,}	
\def\r@fcite#1,{\ifuncit@d{#1}
    \newr@f{#1}%
    \expandafter\gdef\csname r@ftext\number\r@fcount\endcsname%
                     {\message{Reference #1 to be supplied.}%
                      \writer@f#1>>#1 to be supplied.\par}%
 \fi%
 \csname r@fnum#1\endcsname}
\def\ifuncit@d#1{\expandafter\ifx\csname r@fnum#1\endcsname\relax}%
\def\newr@f#1{\global\advance\r@fcount by1%
    \expandafter\xdef\csname r@fnum#1\endcsname{\number\r@fcount}}

\let\r@fis=\refis			
\def\refis#1#2#3\par{\ifuncit@d{#1}
   \newr@f{#1}%
   \w@rnwrite{Reference #1=\number\r@fcount\space is not cited up to now.}\fi%
  \expandafter\gdef\csname r@ftext\csname r@fnum#1\endcsname\endcsname%
  {\writer@f#1>>#2#3\par}}

\def\ignoreuncited{
   \def\refis##1##2##3\par{\ifuncit@d{##1}%
     \else\expandafter\gdef\csname r@ftext\csname
r@fnum##1\endcsname\endcsname%
     {\writer@f##1>>##2##3\par}\fi}}

\def\r@ferr{\endreferences\errmessage{I was expecting to see
\noexpand\endreferences before now;  I have inserted it here.}}
\let\r@ferences=\references
\def\references{\r@ferences\def\endmode{\r@ferr\par\endgroup}}

\let\endr@ferences=\endreferences
\def\endreferences{\r@fcurr=0
  {\loop\ifnum\r@fcurr<\r@fcount
    \advance\r@fcurr by 1\relax\expandafter\r@fis\expandafter{\number\r@fcurr}%
    \csname r@ftext\number\r@fcurr\endcsname%
  \repeat}\gdef\r@ferr{}\endr@ferences}


\let\r@fend=\endpaper\gdef\endpaper{\ifr@ffile
\immediate\write16{Cross References written on []\jobname.REF.}\fi\r@fend}

\catcode`@=12

\citeall\refto		
\citeall\ref		%
\citeall\Ref		%

\catcode`@=11
\newcount\tagnumber\tagnumber=0

\immediate\newwrite\eqnfile
\newif\if@qnfile\@qnfilefalse
\def\write@qn#1{}
\def\writenew@qn#1{}
\def\w@rnwrite#1{\write@qn{#1}\message{#1}}
\def\@rrwrite#1{\write@qn{#1}\errmessage{#1}}

\def\taghead#1{\gdef\t@ghead{#1}\global\tagnumber=0}
\def\t@ghead{}

\expandafter\def\csname @qnnum-3\endcsname
  {{\t@ghead\advance\tagnumber by -3\relax\number\tagnumber}}
\expandafter\def\csname @qnnum-2\endcsname
  {{\t@ghead\advance\tagnumber by -2\relax\number\tagnumber}}
\expandafter\def\csname @qnnum-1\endcsname
  {{\t@ghead\advance\tagnumber by -1\relax\number\tagnumber}}
\expandafter\def\csname @qnnum0\endcsname
  {\t@ghead\number\tagnumber}
\expandafter\def\csname @qnnum+1\endcsname
  {{\t@ghead\advance\tagnumber by 1\relax\number\tagnumber}}
\expandafter\def\csname @qnnum+2\endcsname
  {{\t@ghead\advance\tagnumber by 2\relax\number\tagnumber}}
\expandafter\def\csname @qnnum+3\endcsname
  {{\t@ghead\advance\tagnumber by 3\relax\number\tagnumber}}

\def\equationfile{%
  \@qnfiletrue\immediate\openout\eqnfile=\jobname.eqn%
  \def\write@qn##1{\if@qnfile\immediate\write\eqnfile{##1}\fi}
  \def\writenew@qn##1{\if@qnfile\immediate\write\eqnfile
    {\noexpand\tag{##1} = (\t@ghead\number\tagnumber)}\fi}
}

\def\callall#1{\xdef#1##1{#1{\noexpand\call{##1}}}}
\def\call#1{\each@rg\callr@nge{#1}}

\def\each@rg#1#2{{\let\thecsname=#1\expandafter\first@rg#2,\end,}}
\def\first@rg#1,{\thecsname{#1}\apply@rg}
\def\apply@rg#1,{\ifx\end#1\let\next=\relax%
\else,\thecsname{#1}\let\next=\apply@rg\fi\next}

\def\callr@nge#1{\calldor@nge#1-\end-}
\def\callr@ngeat#1\end-{#1}
\def\calldor@nge#1-#2-{\ifx\end#2\@qneatspace#1 %
  \else\calll@@p{#1}{#2}\callr@ngeat\fi}
\def\calll@@p#1#2{\ifnum#1>#2{\@rrwrite{Equation range #1-#2\space is bad.}
\errhelp{If you call a series of equations by the notation M-N, then M and
N must be integers, and N must be greater than or equal to M.}}\else%
 {\count0=#1\count1=#2\advance\count1
by1\relax\expandafter\@qncall\the\count0,%
  \loop\advance\count0 by1\relax%
    \ifnum\count0<\count1,\expandafter\@qncall\the\count0,%
  \repeat}\fi}

\def\@qneatspace#1#2 {\@qncall#1#2,}
\def\@qncall#1,{\ifunc@lled{#1}{\def\next{#1}\ifx\next\empty\else
  \w@rnwrite{Equation number \noexpand\(>>#1<<) has not been defined yet.}
  >>#1<<\fi}\else\csname @qnnum#1\endcsname\fi}

\let\eqnono=\eqno
\def\eqno(#1){\tag#1}
\def\tag#1$${\eqnono(\displayt@g#1 )$$}

\def\aligntag#1\endaligntag
  $${\gdef\tag##1\\{&(##1 )\cr}\eqalignno{#1\\}$$
  \gdef\tag##1$${\eqnono(\displayt@g##1 )$$}}

\def\eqalignno#1{\displ@y \tabskip\centering
  \halign to\displaywidth{\hfil$\displaystyle{##}$\tabskip\z@skip
    &$\displaystyle{{}##}$\hfil\tabskip\centering
    &\llap{$\displayt@gpar##$}\tabskip\z@skip\crcr
    #1\crcr}}

\def\displayt@gpar(#1){(\displayt@g#1 )}

\def\displayt@g#1 {\rm\ifunc@lled{#1}\global\advance\tagnumber by1
        {\def\next{#1}\ifx\next\empty\else\expandafter
        \xdef\csname @qnnum#1\endcsname{\t@ghead\number\tagnumber}\fi}%
  \writenew@qn{#1}\t@ghead\number\tagnumber\else
        {\edef\next{\t@ghead\number\tagnumber}%
        \expandafter\ifx\csname @qnnum#1\endcsname\next\else
        \w@rnwrite{Equation \noexpand\tag{#1} is a duplicate number.}\fi}%
  \csname @qnnum#1\endcsname\fi}

\def\ifunc@lled#1{\expandafter\ifx\csname @qnnum#1\endcsname\relax}

\let\@qnend=\end\gdef\end{\if@qnfile
\immediate\write16{Equation numbers written on []\jobname.EQN.}\fi\@qnend}

\catcode`@=12


\rightline{NSF-ITP-92-76}
\rightline{TPI-MINN-92/22-T}
\rightline{May  1992}
\vskip.8in
\centerline{\bf Nonfactorizable Amplitudes in Weak Nonleptonic
Decays of Heavy
Mesons}
\bigskip
\centerline{B. Blok
}
\bigskip
\centerline{\sl Institute for Theoretical Physics}
\centerline{\sl University of California at Santa Barbara}
\centerline{\sl Santa Barbara, CA 93106 }
\centerline{\sl and}
\centerline{M. Shifman
}
\bigskip
\centerline{\sl  Theoretical Physics Institute}
\centerline{\sl University of Minnesota}
\centerline{\sl Minnesota, MN 55455}
\bigskip
\abs{ We discuss nonfactorizable $1/N_c$ contributions in the
amplitudes of
non-leptonic exclusive decays of the type $B\rightarrow D\pi$
($N_c$ is the number of colors). In a certain kinematical limit rather
reliable estimates are possible. It is demonstrated that the
nonfactorizable
parts are of the same order as the factorizable
$1/N_c$ parts of the amplitudes, and have the opposite sign.
Thus, an approximate
rule of discarding $1/N_c$ corrections in the nonleptonic exclusive
decays emerges dynamically.
It is shown that this rule is not exact and the degree of compensation
is
different in different channels.
 Our predictions make use of the fact that a key
matrix element that measures  deviations from  factorization can be
 determined using the heavy quark effective theory.
  }
\endpage
\head{1. Introduction}
\par Factorization is used in the weak nonleptonic decays starting
from early
sixties \refto{1}. However, as our knowledge of QCD and weak decay
 phenomenology
deepens, the simple idea of \ref{1}  to factorize two V-A currents
composing
the effective weak Hamiltonian $H_{\rm weak}$
 evolves towards a rather sophisticated
scheme with different aspects characterized by varying degree of
theoretical
 understanding. In the leading order of the $1/N_c$ expansion,
where $N_c$
is the number of colors, factorization of $H_{\rm weak}$ becomes
exact
\refto{2} in the multicolor chromodynamics.\refto{3} This property is
a
common  place now. We will discuss in this paper the factorization
only at
the subleading, i.e. $O(1/N_c)$ level , where it is far
from being obvious. It has been observed in \ref{4} that at this level
the
naive factorization is incompatible with  experimental data. The
 so called rule of discarding  $1/N_c$ corrections has been
formulated
\refto{5}. This rule is one of the most puzzling and theoretically
unclear
elements of the modern procedure of calculating the amplitudes of
weak
 nonleptonic decays (see e. g. \ref{6} for a review).
 Although the rule of discarding  $1/N_c$ works  in  dozens of
channels, nobody succeeded in
finding a dynamical explanation of this approximate regularity
 starting from  first principles, although
some hints exist in the literature (see below).
\par The question we address in this paper is the dynamical origin of
the rule
of discarding $1/N_c$ corrections in the exclusive decays of heavy
mesons in
a certain kinematical limit.
\par To explain the basic idea  consider for
definiteness a particular mode, $\bar B^0_s\rightarrow D^+_s\pi^-$,
although our
 arguments are quite general. The strange quark is chosen as the
spectator in
 order to isolate the  mechanism we are interested in. As a matter of
fact,
if we disregard annihilation graphs (which are presumably
unimportant), a less
exotic mode, $\bar B^0\rightarrow
 D^+\pi^-$, is equally suitable for our purposes.
Below we shall drop the explicit subscript "s"
keeping in mind that we refer to both decays.
\par  The
task is to demonstrate qualitatively the emergence of the  rule
of discarding the $1/N_c$ corrections \refto{5}
under certain kinematic conditions. The cataloguing of the set of
specific
exclusive channels which are of practical interest is postponed till a
separate
publication.
\par To elucidate our approach let us first of all remind that the
relevant
part of $H_{\rm weak}$ includes two operators,
$$\eqalign{&O_1=(\bar c_i\Gamma^{\mu}b^i)(\bar d_j\Gamma_\mu
u^j)\cr
&O_2=(\bar c_i\Gamma^\mu b^j)(\bar d_j\Gamma_\mu u^i)\cr}\eqno
(1)$$
where $\Gamma_{\mu}=\gamma_{\mu}(1+\gamma_5)$.
The effective weak Hamiltonian $H_{\rm weak}$ has the form
$$H_{\rm weak}={G_F\over \sqrt{2}}V_{cb}
V_{du}(c_1O_1+c_2O_2)\eqno (1a)$$
where $c_1\sim 1.12$, $c_2\sim -0.29$ for the case of B-meson
decays
\refto{34} (for a detailed discussion of the numerical values of
$c_{1,2}$
and uncertainties see \ref{51}).
 The first operator in eq. \(1a) is the product of two
color singlet currents, while in the second one the color flow is
twisted. We
will refer to $O_2$ as to the color-twisted operator.
\par In the matrix elements of $O_1$ the factorization procedure just
 prescribes
to factorize the color singlet brackets:
$$<D^+\pi^-\vert O_1\vert \bar B^0> =<\pi^-\vert\bar d
 \Gamma^{\mu}u\vert 0><D^+\vert \bar c
\Gamma_{\mu}b\vert\bar B^0>.\eqno (2)$$
This result is exact in the leading order in $1/N_c$.  Corrections
appear
only at the level of $O(1/N^2_c)$ (see  \ref{7} for a detailed
discussion),
while we limit our discussion to the O$(1/N_c)$ terms. Thus, with our
accuracy, eq. \(2) presents the solution of the problem
for the part of the decay amplitude related to the operator $O_1$.
\par For the color-twisted operator $O_2$ the naive factorization
would yield
$$<D^+\pi^-\vert O_2\vert\bar B^0>=(1/N_c)<\pi^-\vert \bar
d\Gamma^\mu u\vert 0>
<D^+\vert \bar c\Gamma_\mu b\vert\bar B^0>\eqno (3)$$
with the explicit factor $1/N_c$ in the right-hand side. Needless to
say that
eq. \(3) is not justified by $1/N_c$ counting; moreover, this
expression is
not the one welcomed by phenomenology.
\par It has been demonstrated by  analyzing   experimental data
\refto{4,8} that the matrix elements like
$<D^+\pi^-\vert O_2\vert\bar B^0>$
 must contain some nonfactorizable contributions compensating, to a
large
extent, the factorizable part. The effect of compensation seems to
show up
 practically in all channels studied,
 although, certainly, there may remain residual relatively
small uncompensated contributions.
\par The rule of discarding $1/N_c$ corrections crystallized
gradually
and  was formulated as an empirical regularity in \ref{5}. In
 two-particle  D-meson decays it has been confirmed numerically
within
the QCD sum rule approach in \refto{15}. The search for theoretical
reasons
explaining the rule is still a challenge to QCD practitioners. Success in
this
direction,
although takes place but is still very
  limited. \refto{15, 16, 17}
\par In
this paper
we suggest to calculate the nonfactorizable part in $<D^+\pi^-\vert
O_2
\vert\bar B^0>$
in the following limit. Assume that the initial and final heavy quark
masses,
$M_i$ and $M_f$, tend to infinity while their difference $\Delta
M=M_i-M_f$
is held fixed and is of the order of the typical hadronic mass. Under
these
kinematical conditions:
\par (i) The problem can be formulated in terms of the standard
operator
product expansion.
\par (ii) The leading operator that determines
 the nonfactorizable part is of the
$\vec\sigma\vec H$ type (the chromomagnetic moment  , see the
discussion
below). Its matrix element in the $B\rightarrow D$ transition is
unambiguously
fixed within the heavy  quark effective theory.
\par If the actual b and c quarks can be treated in this limit (this
certainly
requires reservations and one should be cautious about the accuracy
of
predictions) the nonfactorizable part
of the amplitude is calculable. We will see below that it
is of the order of the factorizable
 part of the amplitude $<D^+\pi^-\vert O_2\vert\bar B^0>_{\rm f}$
and has the opposite sign.
\par Warning: in the literature the word "factorization" is applied to
quite
different procedures, based on different theoretical assumptions. To
avoid
confusion, one should pay attention to  details. Let us mention a
recent
work \refto{9} done under the assumption that
$M_{i,f}\rightarrow\infty$
and $\Delta M$ scales in the same way as $M_{i,f}$. This seemingly
insignificant
change in kinematics leads to drastically different results. If under
the
assumptions of the present paper the naive factorization of the color
singlet
brackets is likely to be supplemented by the
approximate rule of discarding $1/N_c$
corrections, the assertion of \ref{9} reduces to the naive factorization.
According to \ref{9} the amplitude $<D^+\pi^-\vert O_2\vert\bar
B^0>$
 is exhausted
by eq. \(3) in the limit
 considered by these authors.  Below we shall explain
why our approach is not directly generalizable to include the domain
$\Delta M
\sim M_{i,f}$ analysed in \ref{9}.

\head{2. Operator Product Expansion in $<D^+\pi^-\vert O_2\vert\bar
B^0>$}

\par In this section we develop a version of the QCD sum rule
approach \refto{10}
allowing us to analyze systematically the nonfactorizable parts in the
 amplitudes like $<D\pi\vert O_2\vert\bar B>$
 in the kinematical regime specified in
section 1.
\par The operator $O_2$ can be decomposed in two terms:
$$\eqalign{&O_2=\tilde O_2+(1/N_c)O_1,\cr
&\tilde O_2=2(\bar c \Gamma_\mu t^a b)(\bar d \Gamma^{\mu} t^a
u),\cr}\eqno (4)$$
where $t^a$ stands for the color matrices in the fundamental
representation
of SU(3) (${\rm tr}(t^at^b)= {1\over 2}\delta^{ab}$).
 The naive factorization implies  that the
 $\bar B^0\rightarrow D^+\pi^-$
matrix element of the operator $\tilde O_2$ vanishes. If, on the other
hand,
 $<D^+\pi^-\vert\tilde O_2\vert\bar B^0>\sim -(1/N_c)<D^+\pi^-
\vert O_1\vert\bar B^0>$
the rule of discarding $1/N_c$ corrections will emerge dynamically.
We shall refer below to the amplitude $<D^+\pi^-\vert\tilde
O_2\vert\bar B^0>$
as to the nonfactorizable part of $O_2$.
\par Below we will demonstrate that the color exchange between the
brackets of
the operator $\tilde O_2$ does take place at the level not suppressed
by the
inverse powers of the masses of initial or final heavy mesons
$M_{i,f}$. The only suppression factor, apart from numericals, is
$1/N_c$ $-$
the same as in eq. \(3). This $1/N_c$ factor is in full accordance with
the
 general rules.\refto{3} In order to calculate the nonfactorizable part
of
$O_2$ (color exchange mechanism of Fig.1) let us use a modification
of
 the QCD sum rule approach. \refto{10} We hasten to
add that not all heavy machinery of this method  will be involved,
just a
few aspects that are absolutely simple and transparent. Instead of
the amplitude
 $<D^+\pi^-\vert\tilde O_2\vert\bar B^0>$
 the starting point of our analysis  is
the correlator

$${\cal A}^\beta
=\int d^4x<D\vert T\{\tilde O_2(x),A^\beta (0)\}\vert\bar
B>e^{iqx}\eqno (5)$$
where $A^{\beta}$ is the axial current that annihilates the pion,
$$A^{\beta}=\bar u\gamma^\beta\gamma^5 d.\eqno (5a)$$
The momentum $q$ is an auxiliary momentum flowing through
$A^{\beta}$ which is
assumed to be euclidean. By analytically continuing the amplitude
below the threshold with respect to the light quarks we are able to
formulate
the problem as the short distance operator product expansion  (OPE)
 that is under
theoretical control. Then, by saturating the result by the pion
contribution,
we get an estimate of $<D^+\pi^-\vert \tilde O_2\vert\bar B^0>$ (Fig.
2).
The light quark pair is emitted at the point $x$ and absorbed at the
point zero.
The Fourier transformation,
$\int d^4x e^{iqx}$ (with $q^2=-Q^2 $ being negative) ensures the
validity of
 the operator product expansion in the
 sector of the light degrees of freedom. This
block $-$ the block of the soft gluon emission can be systematically
expanded in
powers of $1/Q^2$:
$$<\bar d(x)\Gamma^{\mu}t^au(x), A^{\beta}(0)>\sim
 {C^{\mu\beta}_{\rho\sigma}\tilde G^{a\rho\sigma}\over
q^2}+...\eqno (5b)$$
Here $\tilde G^a_{\rho\sigma}=(1/2)
\epsilon_{\rho\sigma\alpha\beta}G^{a\alpha\beta}$.
 The leading term
in the operator product expansion comes, as we see from eq. \(5b),
from the
 gluon field operator.  Omitting trivial computational details
let us reproduce the answer:
$$\eqalign{
{\it A}^\beta&=
-2i{1\over 8\pi^2}{q^\alpha q^\beta\over q^2}<D\vert \bar c
\Gamma^{\mu}
t^ag\tilde G^a_{\alpha\mu}b\vert B>\cr
&+\rm{ terms}\quad \rm{ suppressed}\quad \rm{by}
\quad\rm{ powers} \quad\rm{ of}\quad
1/Q^2\cr}\eqno (6)$$
In deriving eq. \(6) we have used the standard external field
technique
( see e.g.
 \ref{11}) and retained only one kinematical structure $-$ namely,
the one
 proportional to $q^{\beta}$ which is relevant to the pion saturation.
\par The expression given in eq. \(6) contains a pion-like pole, and it
is
 tempting to interpret it as exclusively due to pion. Clearly, this is not
 the case since we can not trust eq. \(6) at $Q^2\rightarrow 0$, where
higher
order operators come into play and blow up. Eq. \(6) is valid only
 at sufficiently large $Q^2$. Nevertheless, from what we know about
the
axial current $A^{\beta}$, we can
 say that pion constitutes a sizable fraction of all intermediate states-
there is a pion dominance in $A^{\beta}$ just in the same way as
there is the
$\rho-$meson dominance in the vector current. This observation will
allow us
to extract the pion contribution from eq. \(6).
\par Before we turn to the analysis of the pion contribution let us
notice
that the matrix element in eq. \(6) reduces in our kinematics to a
$\vec\sigma\vec H^at^a$ correlation ($H^a$ is the chromomagnetic
field)
which is readily expressible in terms of an observable quantity $-$
the mass
difference between the vector and pseudoscalar heavy mesons, using
the heavy
quark symmetry. We will return to this point in section 3.
\par One last remark, explaining the role of the kinematical domain
that was
chosen in this paper: $M_{i,f}\rightarrow \infty$, $\Delta M=M_i-
M_f$ fixed,
$\Delta M\le$ few GeV.
 The point is that our approach is based on the standard operator
product expansion for the light quark block. The pion is substituted
by the
auxiliary current at momenta q, and we would like to expand in $Q^{-
2}$, so that
the leading operator enters with the coefficient decreasing like $Q^{-
2}$,
dimension 4 operators enter with coefficient decreasing like $Q^{-4}$,
etc.
Then the series in $Q^{-2}$ can be truncated,
 provided that $Q^2$ is chosen in the GeV$^2$ region, and we are left
with the
leading operator alone (see eq. \(5b)) whose matrix element between
two heavy
mesons is known. The expansion parameter is $\mu^2/Q^2$ where
$\mu$ is a
typical light quark momentum in the hadron.
\par What would happen if, instead, we would like to apply the
present approach
in the kinematical region of \ref{9}? It is easy to see that if $\Delta
M$ is assumed to scale with M the proper expansion orders the
operators
in the expansion in terms of their twist, not dimension, and all
operators
of the same twist, but different dimensions, can become equally
important.
 Indeed, let us compare the operator in eq. \(6) with that having the
two extra
 derivatives, $\tilde G_{\alpha\mu}\rightarrow D_\gamma D_\rho
\tilde G_{\alpha
\mu}$. Its dimension is increased by two units, so that the
corresponding
 coefficient contains an extra $q_{\gamma}q_{\rho}/Q^4$ factor.
Were it
the normal operator product expansion, we would get a relative
suppression
of the order $\mu^2/Q^2$, where $\mu\sim 300$  to $400$
 MeV is a typical momentum
of the light
 quarks or gluons in hadrons. This is not the case, however,
 if $\Delta M$ scales like
M. Indeed, in this kinematics $q_0\sim M_i-M_f$.
 Then, choosing, say, the temporal
 components of $D_\gamma D_\rho$ we see that the relative
contribution of
$D_\gamma D_\rho\tilde G_{\alpha\mu}$ versus $\tilde
G_{\alpha\mu}$ is
$\mu^2q_0^2/Q^4\sim \mu^2(M_i-M_f)^2/Q^4$. The expansion
parameter scales
with M, and we have to go to the domain
 $Q^2\ge \mu (M_i-M_f)$
to ensure  theoretical control over the sum rules.
 It is clear that at
such values of $Q^2$ the pion contribution is negligible. If
we still want to single out the pion contribution by considering the
$Q^2\sim {\rm GeV}^2$ domain we face the problem of summing
the contributions of  operators with arbitrary
number of derivatives, which is clearly beyond our present abilities.
\head{3. The $\vec\sigma\vec H$ correlation.}
\par Let us now examine the crucial matrix element
$<D^+\vert\bar c \Gamma^\mu t^ag\tilde G^a_{\alpha\mu}b\vert\bar
B^0>$
in more detail. According to eq. \(6) it determines the
nonfactorizable part of
$O_2$. Since in the kinematics considered $\delta M\ll M_{i,f}$ the
velocity
of the final heavy quark c  in the rest frame of the initial heavy
quark b
  is a small parameter. In the
limit when the velocity of the final heavy quark $\vec v\rightarrow
0$ and
if we are interested in the dominant contribution to the matrix
element at
hand (i.e. that not
 suppressed by the powers of $ M^{-1}$)
 we can
substitute
  $\Gamma^\mu\rightarrow
\gamma^i\gamma^5$. We obtain
$$\eqalign{
q^{\alpha}<D^+\vert\bar c\Gamma^{\mu}t^ag\tilde
G^a_{\alpha\mu}b\vert
\bar  B^0>&=
q_0<D^+\vert
\bar c \gamma^i\gamma^5t^abg\tilde G^a_{0i}\vert\bar B^0>\cr
&=-q_02M<D^+\vert \bar h_c \vec \sigma\vec H h_b\vert\bar
B^0>\cr}\eqno (7)$$
Here $h_{c,b}$ are the fields of the heavy quark effective field
theory
(HQET) \refto{12}, defined as
$$h_Q={(I+\hat v)\over 2}\exp (im_Qvx)\psi_Q (x).\eqno (7a)$$
and normalized nonrelativistically. The fact that in our kinematics
$(M_iM_f)^{1/2}=M$ is taken into account.
In eq. \(7a)   $Q=c,b$ and  $\psi_Q$ is
 the Dirac spinor that describes the heavy
quark in the Dirac formalism. Note that the fields $h_Q$ are the
two-component spinors, while $\psi$ has four components. The
vector
 $\vec\sigma$ is the spin matrix vector $(\sigma_x, \sigma_y,
\sigma_z)
$ where $\sigma_i$ are the usual Pauli matrices. The field $H^a_i$ is
the
 chromomagnetic field, $H^a_i=\tilde G^a_{0i},\vec H=gt^a\vec H^a$.
Notice that the presence of $\tilde G_{\alpha\mu}$ in the operator
$\bar c \Gamma^\mu t^a g\tilde G_{\alpha\mu}b$ radically changes
the roles of
different components of the $\gamma$  matrices
in comparison with the standard case of the transition driven by
the operator $\bar c\Gamma_\mu b$.
For the standard current $\bar c \Gamma_{\mu} b$ the amplitude of
$B\rightarrow
D$ transition is determined by the
matrix element of the operator $\Gamma_\mu\rightarrow
\gamma_0$; neither the
heavy quark nor the light cloud flip their spins
in this transition (Fig. 3a). As for the operator
$ \bar c\Gamma^{\mu}t^ag\tilde G^a_{\alpha\mu}b$ in the  leading
in M
 approximation one is forced to pick up a component
that flips the spin (and color spin) of the heavy quark. To
compensate for
that the operator $H^a_i$ flips the total angular momentum and the
color spin
of the light cloud (Fig. 3b).
\par For  arbitrary values of the heavy quark velocities
 $\vec v$ the matrix element \(7) must be described
by a new universal function of the Isgur-Wise type \refto{13}:
$$<D^+\vert\bar c \Gamma^\mu t^ag\tilde
G^a_{\alpha\mu}b\vert\bar B^0>
=\sqrt{M_iM_f}
f(y)(v^{i}_\alpha+v^{f}_\alpha).\eqno (8)$$
Here $v_{\alpha}^{i,f}$ are the four-velocities
of the initial and final heavy quarks
 and $y=vv'$ is the recoil. In the limit
$y\rightarrow 1$ $(\vec v\rightarrow 0)$ the function $f$ can be
 normalized in
the same way as the Isgur-Wise function
$\xi (y)$. Recall that the  normalization
theorem of \ref{14}
exactly fixes
the value of the Isgur-Wise function at zero recoil $\xi (y=1)=1$.
Consider now the function $f(y)$.
  The meaning of
$f(y=1)$ is transparent. This parameter measures the degree of the
correlation
of the heavy quark spin with the chromomagnetic field in the light
cloud. Since
in the pseudoscalar mesons the heavy quark spin is $100\%$
 correlated with the total angular momentum of the light cloud, one
can equally
say that $f(y=1)$ measures the degree of correlation of the angular
momentum
of the light cloud with the chromomagnetic field
 $ H^a_i $ in the light cloud. It is physically
clear that this correlation is large, and $f(y=1)$ must be of the order
of the
typical hadronic mass parameter squared ( we mean , of course, the
parameter
associated with the light
degrees of freedom). We will denote
 $$ f(y=1)=m^2_{\sigma H}.\eqno (9)$$
\par Remarkably, the heavy quark effective theory \refto{12, 13}
permits one
 to get a model independent estimate of $m^2_{\sigma H}$. Indeed,
due to
the flavor symmetry we have:
$$f(y=1)=m^2_{\sigma H}=-<\bar B\vert
 \bar h_b\vec\sigma g\vec H^at^ah_b\vert\bar B>.\eqno (10)
$$
The operator in the r.h.s. of
eq. \(10) divided over $2M$
 is the subleading operator in the Hamiltonian of the heavy quark
effective theory. \refto{12,30} Moreover this is the only operator (up
to
 terms
$O(\Lambda^2_{\rm QCD} /M^2))$ that breaks the spin symmetry,
 present in the limit
$M\rightarrow\infty$.
 Therefore, we immediately get the value of its matrix element over
the B-meson.
  Indeed,
 the spin term in the Hamiltonian (in the nonrelativistic notation) is
$$H={\vec\sigma\vec H\over 2M}\eqno (10a)$$
 The spin shifts in the masses of B-mesons are $\Delta
M(B^*)={1\over 4}\Delta$,
$\Delta M(B)=-{3\over 4}\Delta$ where $\Delta =M_{B^*}-M_B$.
Here $M_{B^*}$ is the mass of B$^*$ meson and $M_B$ is the mass of
the B-meson.
 The
 matrix element
$1/(2M)<\bar B\vert \bar h_b\vec\sigma\vec H
 h_b\vert\bar B>$ in eq. \(10) is
 the matrix element of the spin Hamiltonian $H$.
 The matrix element of the latter over the B meson is equal to $-
{3\over 4}
(M_{B^*}-M_B)$. From this we get
$$-{3\over 4}(M^2_{B^*}-M^2_B)=<\bar B
\vert\bar h_b\vec\sigma g\vec H^a t^ah_b\vert\bar B>
\eqno (11)$$
 Equation \(11) is exact in the limit $M\rightarrow \infty$ and is the
analog
of the normalization condition for the Isgur-Wise function
that was discussed above.
\par In order to obtain the numerical value of $m^2_{\sigma H}$ we
shall use
the recent results on the $B^*-B$ mass difference: $M^2_{B^*}-
M^2_B\sim
0.46 \quad {\rm GeV}^2$ \refto{31,32}.
 Substituting the data we arrive at
$$m^2_{\sigma H}\simeq +0.35\,\,\, {\rm GeV}^2.\eqno (12)$$
For convenience of future references let us remind that
$$q^\alpha <D^+\vert\bar c \Gamma^{\mu}t^ag\tilde G_{\alpha\mu}b
\vert\bar B^0>_{y
\rightarrow 1}\sim q_0 2\sqrt{M_iM_f} m^2_{\sigma H}\eqno (13)$$
(see eqs. \(7), \(8), \(10)).
\head{4.  Saturation by the pion.}
 \par We return now to eqs. \(5), \(6)
 intending to saturate the axial current $A^\beta$
 by the pion contribution (pion
dominance). As it was already mentioned, the fact that eq. \(6)
contains the
pole $q^{-2}$ does not mean that the correlation function \(5) is
exactly
saturated by the pion contribution. The $q^{-2}$ term is only the first
term
in the expansion, and higher order terms
 $-$  $O(q^{-4}),O(q^{-6})$, etc. $-$ signal the
presence of higher states, contaminating the correlation function \(5).
 Potentially, the most important contamination is due to the light
axial meson
$a_1$.
\par At this stage it is very difficult to calculate its relative weight,
although some qualitative arguments can be given. We will comment
on this
 point later, when we shall discuss the accuracy of the final result.
Let us
 assume
at first that we can neglect all higher states, keeping in eq. \(5) only
the
 pion contribution.
We shall also neglect small corrections due to nonzero recoil, i.e.
we put $y=1$.
 Then it becomes straightforward  to obtain from eq. \(6)
a prediction for $<D^+\pi^-\vert\tilde O_2\vert\bar B^0>$. This can be
done just by amputating $-f_\pi q^\beta /q^2$:
$$\eqalign{&
<D^+\pi^-\vert\tilde O_2\vert\bar B^0>_{\rm n.f.}
 ={i\over 4\pi^2f_\pi}q_{\alpha}
<D^+\vert \bar c\Gamma^\mu t^ag\tilde G^a_{\alpha\mu}b\vert \bar
B^0>\cr
&={ix\over 4\pi^2f_\pi}q_02Mm^2_{\sigma H},\quad x=1.\cr}\eqno
(14)$$
Here we used eq. \(13).
The subscript "n.f." stands for "nonfactorizable".
 We have introduced in the second line in eq. \(14) a
correction factor $x$ which is equal to 1 if only the pion contribution
is
 retained in eq. \(6). In general $x$ measures the contamination due
to higher
states. We will argue that the the actual value of $x$ is less than 1,
although
presumably close to one.
\par The factorizable contribution ( it will be marked
by the subscript "f" below )  can be read off from eq. \(3),

$$<D^+\pi^-\vert O_2\vert\bar B^0
>_{\rm f}=-{if_\pi\o N_c}q_0 2M.
\eqno (15)$$
As a result the ratio of the nonfactorizable over the factorizable part
 reduces to

$$r={<D^+\pi^-\vert O_2\vert\bar B^0>_{\rm n.f.}\over
 <D^+\pi^-\vert O_2\vert\bar B^0>_{\rm f}}=-N_c{xm^2_{\sigma
H}\over 4\pi^2
f^2_{\pi}}.\eqno (16)$$
We pause here to note that this ratio is $O(N_c^0)$ and $O(M^0)$. As
far as the
 numerical values are concerned, the ratio $r$ clearly depends on the
parameters $x$ and $m^2_{\sigma H}$. If $x\sim 1$ (complete
pion dominance) and $m^2_{\sigma H}
\sim 0.35\quad \rm{GeV}^2$ we obtain
$$r\simeq -1.5.\eqno (17)$$
Of course, the accuracy of this estimate is not very high. There are
two
main  reasons
for it. First, we use the standard operator product expansion keeping
only the leading term. This should be justified in the limit
 $M_{i,f}\rightarrow\infty, \Delta M\sim E_{\pi}$
fixed and small. The question is whether the actual B to D transitions
fall
into this category. In this case $E_\pi\sim 3\quad {\rm GeV}$
 and $\Delta M$ is comparable
to $M_{i,f}$.
\par As has been mentioned previously, the set of relevant operators
in the
problem at hand includes, e.
g.,
$$
\hat D_n=q^{\alpha_1}....q^{\alpha_n}{D_{\alpha_1}...D_{\alpha_n}
\tilde G_{\beta\gamma}
\over Q^{2n+2}}\eqno (18)$$
The contribution of the operator \(18) in the amplitude \(5) is
proportional
to $(E^n\mu^n/Q^{2n})Q^{-2}$ where $\mu$ is a typical hadronic scale
coming
from the matrix element $D_{\alpha_1}... D_{\alpha_n} \tilde
G_{\beta\gamma}$.
 One can expect
that $\mu\sim 0.3$  to $0.4$ GeV, i.e. the characteristic light
quark (gluon)  momentum in hadrons. Then the expansion parameter
 $$\lambda=E_\pi\mu/Q^2\eqno (18a)$$
becomes of order 1 in the kinematic domain relevant to the
 $\bar B\rightarrow D\pi$ transition
($E\sim 3\quad {\rm GeV},\quad Q^2\sim 1\quad {\rm GeV}^2)$.
 Hence, we see that the actual B  decays
 are at the borderline of our formalism. This fact can introduce an
error of
order one in our estimates of\hfill\break
$ <D^+\pi^-\vert O_2\vert\bar B^0>$.
\par The second source of uncertainty is the
uncertainty in the parameter $x$ that measures the degree of
 pion saturation. Let us notice that if the ratio analogous to $r$ in eq.
\(16)
for $a_1$ has the same (negative) sign $-$ which seems unavoidable
$-$
then the sign
of contaminating contributions in eq. \(6) is fixed. It is the same as
for the
pion contribution. Consequently we can predict that $x\le 1$.
\par  It is known that in the correlation function of two axial
currents at $Q^2\sim 1$ GeV$^2$ the pion and $a_1$ contributions
are
approximately equal \refto{10}. The corresponding polarisation
operator is
characterized by the spectral density growing with  $s$ (the
  energy squared) increasing.
On the other hand from eq. \(6) we see that in the correlation
function we
are interested in the spectral density is concentrated at low values of
s.
Otherwise we would not get
 the leading asymptotics $\sim 1/Q^2$. This
means that the ratio of $a_1$ over $\pi$ contributions
must be  significantly smaller
than in the correlation function of two axial currents. In other words
we
expect $x$ to be rather close to 1.
\par Finally, we have another source of uncertainty.
 Recall that in our estimates we neglected
possible recoil effects connected with the nonzero mass difference
between
the actual
 D and B mesons. This is certainly true in the kinematical domain
accepted
 in this paper. In the real world the mass of the D-meson
is $m_D\sim 1.85\quad {\rm GeV}$ and the mass of the B-meson is
$m_B\sim
5.25$ GeV.
 Hence the recoil $y\sim 1.5$. At this
value of $y$ the Isgur-Wise function is of order $\xi(1.5)\sim
0.5$.\refto{35}
 Although
we do not know the evolution of of the function $f(y)$ introduced in
eq. \(8),
it is reasonable to expect that it has approximately the same
dependence on $y$
as the Isgur-Wise function. Thus, although there can be an additional
 correction factor of order 0.5 in each of the amplitudes \(14) and
\(15) ,
it is reasonable to expect that in the ratio
$r$ given by eq. \(16) these correction factors largely cancel each
other.
Then we have only a small uncertainty in $ r$ due to this source.
\par Summarizing,
 it seems justified to say
 that eq. \(17) is a strong indication that the approximate rule of
discarding
the $1/N_c$ terms  emerges dynamically in the
$\bar B\rightarrow D\pi$ transitions.
\head{5.  Remarks on other decays and conclusions.}
 \par Above we concentrated on $\bar B^0\rightarrow D^-\pi^+$
decays .
 Let us now
discuss in brief other decays.
Consider first the B meson decays. All these decays can be split in
two
classes. The first group includes decays whose amplitudes contain a
factor
$(c_1+{c_2\over N_c})$ in the naive factorization approach. The
second
 group includes decays that contain a factor $(c_2+{c_1\over N_c})$
 in the amplitude
or a sum of two terms proportional to each of the above factors
if we use the naive factorization approach. The amplitudes of the
second
class  would contain in our approach the matrix elements like
$ <\pi^-\vert\bar u \tilde G_{\alpha\rho}
\gamma^\rho\gamma^5 b\vert\bar B>$
or $ <\rho^-\vert \bar u \tilde G_{\alpha\rho}\gamma^{\rho}b
\vert\bar B>$.
Unfortunately, we can not connect these matrix elements to any
vacuum
averages in HQET or to any experimentally measured quantities.
Instead,
we need additional approximations to study them (like QCD sum
rules,
 etc).
We shall present the corresponding analysis elsewhere.
\par Let consider the first group
 of decays. Our approach can be easily extended
  to the decays like $\bar B ^0\rightarrow D^+\rho^-$ and $\bar
B^0\rightarrow
D^{+*}\pi^-$.  For the case of
 $\bar B^{0}\rightarrow D^{+}\rho^{-}$ we can
essentially repeat all the arguments of sections 2-4. The main
difference is
that we now shall consider the vector current
 $\bar u\gamma_\mu d$ instead of the axial one.
Using the conservation of the  current $\bar u\gamma_\mu d$
and assuming the $\rho$-meson dominance in the corresponding
polarisation
operator it is easy to extract the amplitude of the decay
$\bar B ^0\rightarrow D^-\rho^+$. We get
for the parameter $r'$ that measures the ratio of the nonfactorizable
and color suppressed factorizable terms in the amplitude of this
decay:
$$r'={{\cal A}_{\rm n.f.}\over {\cal A}_{\rm f}}
\sim -{N_cx'm^2_{\sigma H}\over 4\pi^2f^2_\rho}1.6.
\eqno (19)$$
The factor $1.6$ comes from the ratio ($Q^2+m^2_\rho)/Q^2$ at
$Q^2=1$ GeV$^2$.
It gives the idea of the theoretical uncertainty. The parameter
 $x'$ is a new correction constant that measures the contamination
of the corresponding correlation function by higher resonances;
$x'=1$ if
we assume the complete $\rho$-meson dominance.
 The $\rho$-meson leptonic decay constant
$f_\rho$ is determined as $<0\vert j^V_\mu\vert
\rho>=f_{\rho}m_\rho e_\mu$,
where $e_\mu$ is the $\rho -$ meson polarisation vector
and $j^V_\mu=\bar u\gamma_{\mu} d$ is the vector current.
 We use for $f_\rho$
the standard value $f_{\rho}\simeq 200$ MeV. Assuming $x'\simeq
1$ as in the case of the
axial current  we get
$$r'\sim -1.\eqno (20)$$
In this decay we once again observe the tendency towards
compensation of
factorizable and nonfactorizable parts. Note that we do not have to
introduce
any new matrix elements to study this decay.
\par Consider now the
$\bar B^0\rightarrow
D^{+*}\pi^-$ decay. In this case we can repeat all considerations above
once again and end up with the matrix element $<D^{+*}\vert\bar c
\tilde G_{\alpha \mu}\gamma^\mu b\vert\bar B^0>$. In the limit of
HQET
(in this limit we must make a substitution
 $\Gamma^\mu\rightarrow\gamma_0$) this matrix
element becomes proportional to $<D^{+*}\vert \bar h_c\vec H h_b
\vert\bar B^0>$. The corresponding  amplitude is
$$<D^{+*}\pi^-
\vert \tilde O_2\vert\bar B^0>_{\rm f}={i\over 4\pi^2}
(-\vec q)
<D^{+*}\vert \bar c\vec H b\vert\bar B^0>.\eqno (111)$$
 Using the commutation relations of HQET
it is easy to see that this matrix element reduces to
$$-\vec q<D^{+*}\vert \bar c\vec Hb\vert\bar B^0>\rightarrow(\vec
q\cdot\vec
 \epsilon){1\over 3}m^2_{
\sigma H}.\eqno (21)$$
Here $\epsilon$ is the polarisation vector of $D^*$ meson.
Thus,

$$<D^{+*}\pi^-\vert O_2\vert\bar B^0>_{n.f.}
=i{(\vec q\cdot\vec e)m^2_{\sigma H}\over 12\pi^2f_\pi} 2M\eqno
(22a)$$
while
$$<D^{+*}\pi^-
\vert O_2\vert\bar B^0>_{\rm f}=-i(\vec q\cdot\vec e){f_\pi \over
N_c} 2M.
\eqno (22)$$

Notice that with our sign conventions $<D^{+*}\vert
\bar c \gamma^{\mu}\gamma^5b\vert\bar B^0>_{\vert_{y=1}}=-
2M\epsilon^\mu$
i.e. we have an extra minus sign as compared to \ref{13}.
The  ratio of nonfactorizable and factorizable parts of the color
suppressed amplitude is
$$r^{''}=-{N_cm^2_{\sigma H}\over 12\pi^2 f^2_\pi}\sim -0.5.\eqno (23)$$
  The ratio $r''=r/3$, i.e. it is three times smaller, although  has
the same sign as the corresponding ratio $r$ for the $\bar
B\rightarrow D\pi$
decay. This result underlines the dynamical nature of the
rule of discarding the  $1/N_c$ corrections. Although the
nonperturbative
color exchange
effects tend to cancel the $1/N_c$ part of the factorized amplitude
the
degree of this cancellation can be different in different channels.
The relation
$$Br(\bar B^0\rightarrow D^+\pi^-)=Br(\bar B^0\rightarrow
D^{*+}\pi^-)\eqno (73)
$$
 present in the
naive factorization approach (neglecting very small kinematical
corrections) is now broken. In particular, we expect
the ratio
$$R={Br
(\bar B^0\rightarrow D^+\pi^-)\over Br(\bar B^0
\rightarrow D^{*+}\pi^-)}>1\eqno (74)$$
even in the heavy quark limit. Substituting the estimates for $r$ and
$r'$
we obtain $$R\sim 1.2.\eqno (75)$$
 This result can be compared with the value $R\sim 1.03$
in the Bauer-Stech-Wirbel model
(see e. g. \ref{40} for a recent review)
 and in the naive factorization approach. (
R slightly differs from 1 in these models only
 due to purely kinematical reasons). Of course we
must look very cautiously at the prediction \(75) due to the
uncertainties in our approach discussed above. Nevertheless,
it would be very interesting to measure $R$ experimentally
and look for its possible enhancement due to unequal cancellation
of $1/N_c$ corrections in these two  channels. Unfortunately, the
present
status of the experimental data does not allow any definite
conclusion
on the deviation of R from unity.
\par  It is interesting to extrapolate our results to the case
of $D\rightarrow K\pi$ decays.
 For these decays  the OPE based method seems even more suitable
 than for the B-meson case, since the expansion parameter
$\lambda$ that was introduced above ( see eq. \(18a) )
 is $\sim 0.4$. The problem is that
we can not use the heavy quark effective theory to find the relevant
matrix
elements in this case. Indeed, the relevant matrix element
is
$$
M=q^\alpha <K\vert \bar s \gamma^{\mu}\gamma^5\tilde
G_{\alpha\mu}c\vert D>
\eqno (231)$$
and it is not known.
Just for the purpose of orientation let us assume, however, that the
 $\vec\sigma\vec H$ correlation in eq. \(231) is the same as for the
 $B\rightarrow D$ transition. Then the nonfactorizable part of
$(\bar s_i\Gamma_\mu c^j)(\bar u_j\Gamma^\mu d^i)$ in the
$D^0\rightarrow
K^-\pi^+$ transition can be read off from eq. \(6), with an obvious
 substitution $q_02M\rightarrow q^{\alpha}f^+ (p_D+p_K)_{\alpha}$,
where
$f^+$ is the transition formfactor for the matrix element \(231), that
is analogous to the standard formfactor $f^+_K$ for semileptonic
transitions
$D\rightarrow K$.
As a result, the nonfactorizable part of the amplitude
 ${\cal A}(D^0\rightarrow K^-\pi^+)$ reduces to
$$<K^-\pi^+\vert O_2\vert D^0>_{n.f.}\sim i\{{f^+ \over
4\pi^2f_\pi}(m^2_D-m^2_K)
m^2_{\sigma H}\}.\eqno (734)$$
The number in the figure brackets is to be compared with the
parameter ${\cal M}_2$
introduced in \ref{15}. Assuming $f^+=1$
we  get for the expression in figure brackets the
value $\sim 0.22 $ GeV$^3$ (cf. ${\cal M}_2=0.11$ GeV$^3$ according
to \Ref{15}).
 This estimate can be improved
if we take into account the  nonzero recoil
in these decays by assuming that the matrix element
\(231) depends on the the recoil $y$ roughly speaking in the
same way as the semileptonic transition formfactor
$<K\vert\bar s \gamma_\mu c\vert D>=f^+_K(p_1+p_2)_\mu$.
  For the latter
we know from \ref{33}
 that $f^+_K\sim 0.5$. On the other hand if
K and D had the same masses, i. e. if we
neglected the recoil effects we would
 have $f^+_K$=1 due to conservation of the
vector current. Keeping this in mind
  we assume that the recoil effects lead to the
additional factor $\sim 0.5$ in the expression for the matrix element
\(231):
$f^+\sim 0.5$.  Then we  get for the nonfactorizable
part of the amplitude in the $D^0\rightarrow K^-\pi^+$ transition:
$$-i<K^-\pi^+\vert O_2\vert D^0>_{\rm n.f.}
\sim +{0.5 m^2_{\sigma H}\over 4\pi^2f_\pi}(m^2_D-m^2_K)\sim
+0.1\quad
{\rm GeV}^3\eqno (24)$$
\par Of course, one should not pay too much attention to the literal
coincidence
of the numerical predictions for the nonfactorizable
parts in this work versus \ref{15} . First, the estimates here for the
$B\rightarrow D$ transition have an intrinsic uncertainty (see
above). Second,
extrapolation from $B\rightarrow D$ to $D\rightarrow K$ transitions
is
semiquantitative at best. Finally, in the analysis of \ref{15} we see
now some
elements that can be rectified giving a somewhat smaller values of
the
 nonfactorizable parts
 ${\cal M}_{1,2,3}$ introduced in that paper (although within the
error
bars).
\par Summarizing, we have shown that there are good reasons to
believe that
the rule of discarding $1/N_c$ corrections can approximately take
place in
heavy meson decays due to dynamical reasons. It is not exact
however, and
in high precision measurements one can expect to observe deviations
from
 this rule. Previously, dynamical indications on the emergence of this
rule were found  in different kinematical domains
\refto{15,16,17}. The task of bringing all
seemingly unrelated analyses together to obtain a unified
picture remains a challenge. It is desirable to establish not only when
the
 rule of discarding $1/N_c$ is valid, but also when we begin
 to find some deviations thus underlining its dynamical nature.
\par This work has been started during the Workshop on Heavy
Quarks
 in ITP, Santa Barbara, January 1992.
 One of the authors (M.S.) is grateful to the
 organisers of the Workshop and the ITP staff for
hospitality and financial support. He is also grateful to A. Vainshtein
for
useful discussions.
\endpage
\references

\refis{1} J. Schwinger, Phys. Rev. Lett., 12 (1964) 630.

\refis{2} A. A. Migdal, 1982, unpublished.

\refis{3} G.'t Hooft, Nucl. Phys., B72 (1974) 461;\hfill\break
          E. Witten, Nucl. Phys., B149 (1979) 285.

\refis{4} M. Bauer, B. Stech and M. Wirbel, Z. Phys., C34 (1987) 103.

\refis{5} A. J. Buras, J.-M. Gerard and R. Ruckl, Nucl. Phys., B268
(1986) 16.

\refis{6} M. Shifman, Int. J. Mod. Phys., A3 (1988) 2769.

\refis{7} M. Shifman, ed., Vacuum structure and QCD sum rules.
Reprint volume,
North Hollands, 1992.

\refis{8} D. Bortoletto et al, (CLEO collab.), preprint CLNS 91/1102,
1991.

\refis{9} M. Dugan and B. Grinstein, Phys. Lett., B255 (1991) 583.

\refis{10} M. Shifman, A. Vainshtein and V. Zakharov, Nucl. Phys.,
B147
 (1979) 385, 448.

\refis{17} M. Shifman, preprint TPI-MINN-91/46-T, 1992 (submitted
to Nucl.
 Phys. B).

\refis{11}  V. A. Novikov, M. A. Shifman, A. I. Vainshtein and V. I.
Zakharov,
  Fortschr. Phys., 32 (1984) 585.

\refis{12} E. Eichten and B. Hill, Phys. Lett., B234 (1990)
511;\hfill\break
           H. Georgi, Phys. Lett., B240 (1990) 447.

\refis{13} N. Isgur and M. Wise, Phys. Lett. B232 (1989) 113;
           Phys. Lett., B237 (1990) 527.

\refis{14} M. Voloshin and M. Shifman, Yad. Fiz., 47 (1988) 801 [Sov.
Journ. of Nucl. Phys., 47
 (1988) 511].

\refis{15} B. Blok and M. A. Shifman, Yad. Fiz., 45 (1987), 211, 478,
841 [Sov. Journ. Nucl. Phys. 45 (1987) 135; 301; 522].

\refis{16} A. J. Buras and J.-M. Gerard, Nucl. Phys., B264 (1986) 371.

\refis{30} J. M. Flynn and B. Hill, Phys. Lett., B264, (1991) 173.

\refis{31} J. Lee-Franzini et al, Phys. Rev. Lett., 65 (1990) 2947.

\refis{32} P. S. Akerib et al, Phys. Rev. Lett., 67 (1991) 692.

\refis{33} T. Aliev, V. Eletsky and I. Kogan, Yad. Fiz., 40 (1984) 823 [Sov.
Journ. Nucl. Phys. 40 (1984) 527].

\refis{34} G. Altarelli and L. Maiani, Phys. Lett., B52, 351
(1974);\hfill\break
           M. K. Gaillard and B. W. Lee, Phys. Rev. Lett., 33 (1974) 108.

\refis{35} D. Bortoletto and S. Stone, Phys. Rev. Lett., 65 (1990) 2951.

\refis{40} M. Neubert, V. Rieckert, B. Stech and Q. P. Xu, preprint
           HD-THEP-91-28, to be published in the forthcoming Review
Volume
           on Heavy Flavors.

\refis{51} R. Ruckl, Weak decays of heavy flavors,
Habilitationsschrift,
           Munich University, 1983.

\endreferences
\endpage
\centerline{\bf Figure Captions.}
\bigskip
{\bf Fig.1:} Color exchange amplitudes in the weak exclusive decays.

{\bf Fig.2:} Pion saturation. $A^\beta$ is an auxiliary axial current.

{\bf Fig.3a:} B$\rightarrow$ D transition for the current $\bar
c\Gamma_\mu b$.

{\bf Fig.3b:} B$\rightarrow$ D transition for the current $\bar
c\Gamma^\mu t^a
g\tilde G^a_{\alpha\mu}b $.

\endpage
\endpaper
\end\bye